\documentclass[conference,comsoc]{IEEEtran}
\IEEEoverridecommandlockouts

\usepackage{verbatim,graphicx,epstopdf}
\usepackage[cmex10]{amsmath}
\usepackage[T1]{fontenc}
\usepackage{times}
\usepackage{ifthen}
\usepackage{latexsym}
\usepackage{afterpage}
\usepackage{verbatim}
\usepackage{url}
\usepackage{tabularx}
\usepackage{multirow}

\fontfamily{ptm}\selectfont

\newboolean{draftmode}
\setboolean{draftmode}{false}
\newcommand{\tabincell}[2]{\begin{tabular}{@{}#1@{}}#2\end{tabular}}
\begin{document}

\title{Detection and Decoding for 2D Magnetic Recording
Channels with 2D Intersymbol Interference \vspace{-0.15in}}
\author{\IEEEauthorblockN{Jiyang Yu, Michael Carosino, Krishnamoorthy Sivakumar, Benjamin J. Belzer}
\IEEEauthorblockA{School of Electrical Engineering and Computer Science \\
Washington State University, Pullman, WA, 99164-2752\\
Email: \{jyu, mcarosin, siva, belzer\}@eecs.wsu.edu }
\and
\IEEEauthorblockN{Yiming Chen}
\IEEEauthorblockA{Western Digital Corporation, \\
Irvine, CA, 92612, USA \\
Email: yiming.chen@wdc.com }}

\maketitle

\begin{abstract}
This paper considers iterative detection and decoding on the concatenated communication channel consisting of a two-dimensional magnetic recording (TDMR) channel modeled by the four-grain rectangular discrete grain model (DGM) proposed by Kavcic et.\ al., followed by a two-dimensional intersymbol interference (2D-ISI) channel modeled by linear convolution of the DGM model's output with a finite-extent 2D blurring mask followed by addition of white Gaussian noise. An iterative detection and decoding scheme combines TDMR detection, 2D-ISI detection, and soft-in/soft-out (SISO) channel decoding in a structure with two iteration loops. In the first loop, the 2D-ISI channel detector exchanges log-likelihood ratios (LLRs) with the TDMR detector. In the second loop, the TDMR detector exchanges LLRs with a serially concatenated convolutional code (SCCC) decoder.
\begin{comment}
 in order to form an improved estimate of the original coded bits written to the magnetic media. The 2D-ISI detector is based on prior work by the last three authors. 

In the first loop, the TDMR detector computes an average state transition probability gamma by averaging the conditional channel PDF $p(y \mid u, s, s')$ over the 16-valued a-priori probability $P(y)$ received from the 2D-ISI detector. This method demonstrates convergence to low error probabilities when integrated into the full double loop detector. 
\end{comment}
Simulation results for the concatenated TDMR and $2 \times 2$ averaging mask ISI channel with 10 dB SNR show that densities of 0.48 user bits per grain and above can be achieved, corresponding to an areal density of about 
9.6 Terabits/$\mathrm{in^2}$, over the entire range of grain probabilities in the TDMR model. 
\begin{comment}
the channel coding rate (which is one half of the user bit density) required to achieve a bit error rate (BER) of $10^{-5}$ is, on average, only about 7\% lower than the rate on the TDMR channel alone, for all values of the probability $P_2$ of two-bit grains. 
This paper also briefly summarizes a novel iterative multi-row TDMR detector design reported in an Allerton Conference 2013 accepted paper; this TDMR detector can achieve code rate improvements of up to 12\% compared to a previous TDMR detector for the 4-grain DGM published by Pan, et.\ al.\ (IEEE Trans. Mag., 2011).
\end{comment}
\end{abstract}


\section{Introduction}
\label{sec:intro}
Industry is approaching the data storage density limit of 
magnetic disk drives that write data on one-dimensional tracks.
Alternative technologies such as heat-assisted-magnetic-recording (HAMR) 
and bit patterned media recording (BPM) are under active investigation. 
One drawback of most of these techniques is that they require a radical 
redesign of the recording medium \cite{Roger}. Moreover, it is uncertain 
whether they will come on line quickly enough to prevent a plateau in 
magnetic disk storage density in the near to medium term.

This paper considers detection and coding techniques for an alternate 
approach proposed in \cite{Roger} called two-dimensional magnetic recording 
(TDMR), wherein bits are read and written in two dimensions on conventional 
magnetic hard disks. These disks have magnetic grains of different sizes 
packed randomly onto the disk surface. In TDMR, information bits are channel
coded to a density of up to two bits per magnetic grain, and written by a 
special shingled write process that enables high density recording. 
A key problem is that a given magnetic grain retains the polarization of 
the last bit written on it. Hence, if a grain is large enough to contain 
two bit centers, the older bits will be overwritten by the latest one.

A relatively simple 2D TDMR channel model
is the four-grain rectangular discrete-grain model (DGM) introduced
in \cite{Kavcic}, wherein four different grain types are constructed from
one, two, or four small square tiles.
In \cite{Kavcic}, upper and lower bounds for the channel capacity of this
model are derived showing a potential density of 0.6 user bits per grain. 
For a typical media grain density of 20 Teragrains/$\mathrm{in^2}$, this 
corresponds to about 12 Terabits/$\mathrm{in^2}$. This is more than an 
order of magnitude improvement over current hard disk drives, which exceed 
densities of 500 Gigabits/$\mathrm{in^2}$ \cite{LuPan-jour}.

Coding and detection for the four-grain DGM is considered
in a previous paper by Pan, Ryan, et.\ al.\ \cite{LuPan-jour}.
They construct a BCJR \cite{bcjr} detection algorithm that scans
the input image one row (track) at a time. A 16-state trellis
is constructed as the Cartesian product of media states (that capture 
transitions between different grain geometries during a one 
tile move along a row) and data states (that capture the grain overwrite 
effect). It is shown that the number of states can be reduced to six 
by combining equivalent states. After one forward-backward pass through
each row of the input image, the TDMR detector passes soft information
in the form of log-likelihood ratios (LLRs) to a rate 1/4 serially 
concatenated convolutional code (SCCC) with puncturing, which decodes the 
data at the highest rate that achieves a BER of $10^{-5}$ 
(corresponding to the highest possible user bit density). No iteration 
between the TDMR detector and SCCC is done in \cite{LuPan-jour}, although 
the possibility is mentioned.

A two-row BCJR detector for the four-grain TDMR channel model has been 
proposed recently \cite{Carosino}. It considers the outputs of 
two rows and two columns resulting from bits written on three input rows. 
Moreover, soft decision feedback of grain state information from previously 
processed rows is used to aid the estimation of bits on current rows. 
Finally, the TDMR detector and SCCC decoder iteratively exchange LLRs before 
a final estimate of the bits is obtained. This two-row BCJR detector 
increases the code rate by up to 12\% over \cite{LuPan-jour}. 

This paper considers iterative detection and decoding on the concatenated 
communication channel consisting of the four-grain DGM (TDMR channel) model, 
followed by a 2D-ISI channel 
modeled by linear convolution of the DGM model's output with a finite-extent 
2D blurring mask followed by additive white Gaussian noise (AWGN). 
We propose an iterative detection and decoding scheme that combines TDMR 
detection, 2D-ISI detection, and soft-in/soft-out (SISO) channel decoding in 
a structure with two iteration loops. The 2D-ISI detector is based on prior 
work by the last three authors \cite{JRC,JZZ}. 

The novel contributions of this paper are as follows: \\
(1) Simulation of the complete two-dimensional magnetic read-write channel, 
incorporating the error correction coding, grain over-write effects, and 
2D-ISI; \\
(2) A novel iterative scheme consisting of a double loop structure with 
exchange of soft information between the constituent blocks based on the 
``turbo principle;'' \\ 
(3) System parameter optimization using an EXIT chart technique \cite{tbrink}; \\
(4) For the concatenated TDMR and $2 \times 2$ averaging mask ISI channel 
with an SNR of 10 dB (respectively 9 dB), the channel coding rate (which is 
one half of the user bit density in the modeled scenario) required to 
achieve a BER of $10^{-5}$ is shown to be, on average, only 
about 7\% (respectively 10\%) lower than the rate on the TDMR channel alone, 
for all values of the probability $P_2$ of two-bit grains considered. For
the 10 dB SNR case, densities of 0.48 user bits per 
grain and above can be achieved, corresponding to an areal density of about 
9.6 Terabits/$\mathrm{in^2}$.

This paper is organized as follows. Section~\ref{sec:model_arch}
summarizes the four grain DGM. 
Section~\ref{sec:det/dec} provides an overview 
of the system architecture, explaining the double loop structure. 
Optimization of system parameters using EXIT chart techniques is 
presented in section~\ref{sec:EXIT}. 
Section~\ref{sec:sims} provides simulation results, and 
section~\ref{sec:conc} concludes the paper.

\section{System Model}
\label{sec:model_arch}
\begin{figure*}[tbh]
  \begin{center}
  \includegraphics[width = 6.0in]{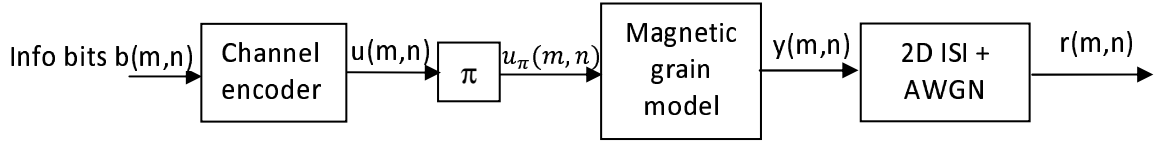}
\caption{Block diagram of write process}
\label{fig:Tx_overall_block}
  \end{center}
\end{figure*}

Figure~\ref{fig:Tx_overall_block} shows a block diagram of the 
write process (transmission model).
A block of user information bits, denoted as $b(m, n)$ in the block diagram, 
is encoded by a rate 1/4 SCCC consisting of an eight state rate 1/2 outer 
non-recursive convolutional code (NRCC) with generator matrix 
$G_{1}(X) = [1 + X, 1 + X + X^{3}]$, 
followed by an interleaver $\pi_{1}$, 
followed by an inner eight state 
recursive systematic convolutional code (RCC) with generator matrix 
$G_{2}(X) = [1, (1 + X + X^{3})/(1 + X)]$, 
followed by a second interleaver $\pi_{2}$. 
Code rates greater than (respectively, less than) 1/4 are achieved by 
puncturing (respectively, repeating) randomly selected output bits from the 
inner encoder. To compare with the result of \cite{LuPan-jour}, the SCCC code in this paper is identical to the one in \cite{LuPan-jour}. See \cite{LuPan-jour,Carosino} for more details about the SCCC encoder and 
decoder.   

The code word bits $u(m,n)$ are level shifted so that 
$(0,1) \rightarrow (-1,1)$, and then are interleaved again to remove the 
dependency between code word bits, and then are written onto the 
TDMR channel with 
over-write property. The TDMR corrupted block of bits is denoted as 
$y(m, n)$, with $y(m, n) \in \{-1,1\}$. 

\begin{figure}[htb]
  \begin{center}
\includegraphics[width = 1.0in]{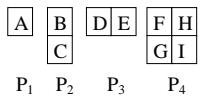}
    \caption{Four-grain rectangular discrete grain model from \cite{Kavcic}.}
    \label{fig:DGMmodel}
  \end{center}
\end{figure}

The TDMR channel consists of a rectangular array (image) of unit-sized 
pixels, where one coded bit is written on each pixel. The image is a 
(random) combination of four distinct types of rectangular grains. 
Relative to the 
unit size their sizes are $1\times 1$, $2\times 1$, $1\times 2$ and 
$2\times 2$ as shown in Fig.~\ref{fig:DGMmodel}.
The four grain types occur with probabilities  
$P_{1}$, $P_{2}$, $P_{3}$, and $P_{4}$, respectively. We assume 
that the average number of coded bits per grain is 2, i.e. 
$1 P_{1} + 2 P_{2} + 2 P_{3} + 4 P_{4} = 2$.  
We also assume that the TDMR channel satisfies an isotropy condition so that 
type 2 and type 3 grains are equally likely, i.e., 
$P_{2} = P_{3}$. 
Based on the above assumptions, all of the four probabilities can be 
computed for a given value of $P_{2} = P_{3}$. To model the write process, the 
TDMR image pixels are given the values $\pm 1$ (equiprobable) as coded bits 
in a row-by-row raster scan order. The multi-bit grains can have only one 
polarity which is determined by the sign of the last bit written on them. 
This property is known as the over-write property of TDMR channel.

\begin{figure}[htb]
  \begin{center}
\includegraphics[width=2.5in]{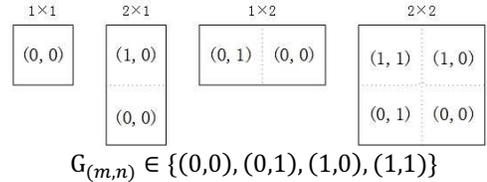}
    \caption{Quadrant notation defined in \cite{Kavcic}.}
    \label{fig:Quadrant}
  \end{center}
\end{figure}

In \cite{Kavcic}, the quadrant notation was introduced to model the 
over-write process (see Fig.~\ref{fig:Quadrant}).
The quadrant notation $G$ is defined as follows:
\begin{equation}
\begin{split}
G_{(m,n)}=\left\{\begin{matrix}
(0,0), \quad &\text{if on grain } A, C, E, I\\
(1,0), \quad &\text{if on grain } B, H\\
(0,1), \quad &\text{if on grain } D, G\\
(1,1), \quad &\text{if on grain } F.\\
\end{matrix}\right.
\label{eq:y_u_relation}
\end{split}
\end{equation}
In (\ref{eq:y_u_relation}), the grain labels A-I refer to 
Fig.~\ref{fig:DGMmodel}.
Using the quadrant notation, we can write the relation between $y(m, n)$ and 
the relevant interleaved coded bit $u_{\pi}(k,l)$ as follows:
\begin{equation}
%\begin{split}
y(m,n)= u_{\pi}((m,n)+G_{(m,n)}),
\label{eq:y_u_relation_new}
%\end{split}
\end{equation}
which models the over-write property by imposing an appropriate spatial 
shift. 

Finally, the cross-track and down-track reading process 
will introduce 2D intersymbol interference (ISI). The TDMR channel with 
arbitrary 2D-ISI can be modeled as follows \cite{Kavcic}: 
\begin{equation}
%\begin{split}
r(m,n)=\sum_{k,l}h(k,l)y(m-k,n-l)+w(m,n),
\label{eq:2DISIequ}
%\end{split}
\end{equation} 
where \textbf{h} is a 2D read-head impulse response, \textbf{w} is the
discrete AWGN field, and \textbf{y} is the actual bits on the 
magnetic grains as given by \eqref{eq:y_u_relation_new}. 
In this paper, we consider the case in which the 2D 
ISI mask \textbf{h} is a $2 \times 2$ averaging mask; i.e., $h(k, l) = 
0.25$ for $k \in \{0, 1\}$ and $l \in \{0, 1\}$, and 0 elsewhere.  
This means the current 
pixel receives adjacent-pixel interference with the same magnitude as the 
current bit, making this \textbf{h} one of the most
difficult $2 \times 2$ masks to equalize.  Since the mask \textbf{h} operates on each unit pixel 
value $y(m,n)$, and the $y(m,n)$ are correlated by the TDMR write model,
(\ref{eq:y_u_relation_new}) and (\ref{eq:2DISIequ}) incorporate a simple
model of grain- and data-dependent 2D-ISI.

The noise level can be quantified using a signal-to-noise ratio (SNR) 
defined as follows \cite{JRC}:
\begin{equation}
%\begin{split}
\text{SNR} = 10\log_{10}\left ( \frac{\text{Var}\left [ \mathbf{y}* \mathbf{h} 
\right ]}{\sigma_{w}^{2}} \right ),
\label{eq:SNR def}
%\end{split}
\end{equation}
where $*$ denotes the 2D convolution in \eqref{eq:2DISIequ} and 
$\sigma_w^2$ is the variance of the AWGN $w(m, n)$. 
%Therefore the variance of the noise can be computed for a given value of SNR.

\section{Combined Detector and Decoder}
\label{sec:det/dec}
\begin{figure*}[tb]
  \begin{center}
  \includegraphics[width=6.0in]{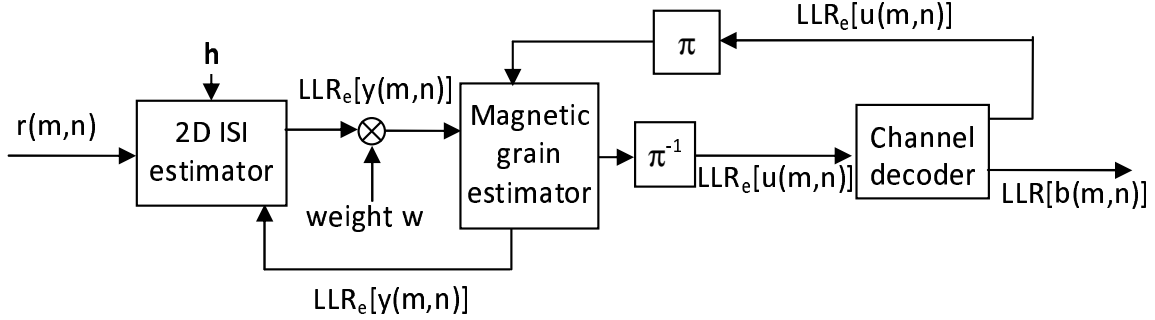}
\caption{Block diagram of read process}
\label{fig:Rx_overall_block}
  \end{center}
\end{figure*}

%
%\subsection{2D-ISI equalization and TDMR detection}
%\label{subsec:2DISI/TDMR}
%
Figure~\ref{fig:Rx_overall_block} depicts the overall block diagram of the 
read process (receiver). 
The receiver consists of two loops: the loop between 2D-ISI equalizer and 
TDMR detector and the loop between TDMR detector and SCCC channel decoder. 
These loops are sometimes referred to as ``outer loops.'' Note that the 
SCCC decoder itself consists of a pair of modules corresponding to the 
constituent encoders described in section \ref{sec:model_arch}. Iterations 
between these decoders within the SCCC decoder are referred to as 
``inner loops.'' 

%For each outer loops, 6 inner loops is executed. 
The received image is first sent into the 2D-ISI equalizer to mitigate 
the effect of 2D-ISI and additive noise and obtain an estimate of the TDMR 
bits $y(m, n)$. The row-column equalization algorithm using joint extrinsic 
information proposed in \cite{JRC} is employed here. This equalizer 
produces soft-output for a block of $2 \times 2$ pixels, resulting in 
16 probabilities for the 16 possible configurations. These probabilities 
are often represented in the log-domain as LLRs. 
The output LLR from the 2D-ISI equalizer is defined as follows: 
\begin{equation}
%\begin{split}
LLR(\mathbf{y}_{k} = \mathbf{m}) = \log 
\left ( \frac{P(\mathbf{y}_{k}=\mathbf{m})}{P(\mathbf{y}_{k}=\mathbf{-1})} 
\right ),
\label{eq:LLR def}
%\end{split}
\end{equation}
where $\mathbf{y}_{k} = \mathbf{m}$ ranges over the 16 configurations of 
$2 \times 2$ images located at position $k$. 
Note that since we serially process the image in a row-by-row order, the row 
position is fixed and the index $k$ merely represents the column position on 
that row. 

The 2D-ISI equalizer output is processed by the TDMR detector's
BCJR algorithm, as described in \cite{Carosino}. 
The TDMR detector estimates LLRs
for coded bits $u(m,n)$ which are deinterleaved and sent to the 
SCCC decoder, after subtraction
of previous input LLRs received from the SCCC.
The SCCC decoder computes LLR estimates of $u(m,n)$, 
which are interleaved 
and fed back to the TDMR detector, after subtraction of SCCC input
LLRs. This loop is repeated several times,
and then the TDMR detector estimates LLRs for $y(m,n)$ which are 
sent back to the 2D-ISI equalizer. The whole process is repeated 
until the SCCC 
decoding converges.

In the following, we describe some details about interfacing the 
2D-ISI equalizer with the TDMR detector and the SCCC decoder. 

\begin{figure}[h]
  \begin{center}
  \includegraphics[width=3.5in]{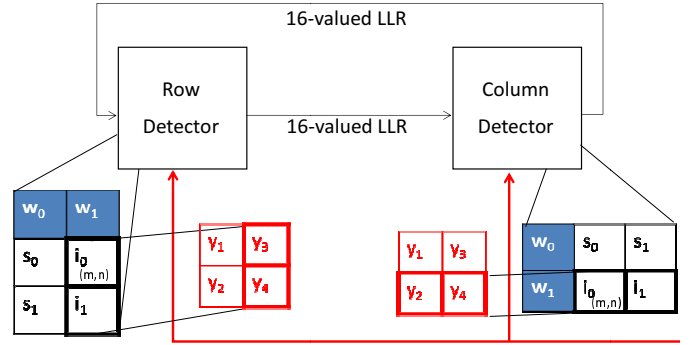}
\caption{Block diagram of 2D-ISI equalizer}
\label{fig:2DISI}
  \end{center}
\end{figure}

Figure~\ref{fig:2DISI} shows the 2D-ISI equalizer structure, based
on that in \cite{JRC}. 
The row detector scans the image in a row-by-row order, whereas the column 
detector scans it column-by-column. In the row/column equalizer, for a given 
image position, the current pixel is denoted as $(m, n)$. The current state 
$s$ consists of two pixels denoted $s_0$ and $s_1$ and the input 
$i$ consists of two pixels denoted $i_0$ and $i_1$. The previous state $s'$ 
consists of the pixels denoted $s_0'$ and $s_1'$. The $2 \times 2$ block of 
pixels $y_1, y_2, y_3, y_4$ are estimates passed on by the TDMR detector and 
serve as extrinsic information for the inputs in the row/column equalizer. 
The pixels denoted $w_0$ 
and $w_1$ are feedback pixels from previous scan direction (e.g., column 
direction for row detector). 
Specifically, in the 2D-ISI BCJR algorithm the $\gamma$ computation is implemented as 
follows: 
\begin{equation}
\begin{split}
\gamma_{\mathbf{i}}\left ( \mathbf{r}_{k},\mathbf{s'},\mathbf{s} \right ) = 
&p'\left ( \mathbf{r}_{k} \mid \mathbf{y}_{k}=\mathbf{i},S_{k}=\mathbf{s},S_{k-1}=\mathbf{s'}\right ) \\
& \times P\left ( \mathbf{y}_{k} \mid \mathbf{s},\mathbf{s}' \right )\times P(\mathbf{s} \mid \mathbf{s'})\\
& \times \hspace{-12pt} \sum_{y(s_0),y(s_1)} \hspace{-12pt} P\left ( \mathbf{y}_{k} \mid \mathbf{\widetilde{y}}_{k} \right ) \times \hspace{-12pt} \sum_{y(s_0),y(s_1)} \hspace{-12pt} P\left ( \mathbf{\mathbf{y}_{k} \mid \widehat{y}}_{k} \right ).
\label{eq:2DISI gamma}
\end{split}
\end{equation}

In \eqref{eq:2DISI gamma}, the factor 
$P\left ( \mathbf{y}_{k} \mid \mathbf{s},\mathbf{s}' \right )$ is either 
0 or 1 based on trellis structure, and $P(\mathbf{s} \mid \mathbf{s'})$ is 
a constant for all valid state transitions (and zero otherwise).
$P\left ( \mathbf{y}_{k} \mid \mathbf{\widetilde{y}}_{k} \right )$ 
is the extrinsic feedback probability from the other (row or column) detector. 
$P\left (\mathbf{y}_{k} \mid \mathbf{\widehat{y}}_{k} \right )$ is 
the 16-valued feedback probability from the TDMR detector. 
Since the same $2 \times 2$ $\mathbf{y}$ vectors are 
used in both row and column detector, marginalization is needed  
in each equalizer. For example, as shown in Fig.~\ref{fig:2DISI},
to compute the
{\em a priori} joint input probability $P(i_0,i_1)$ for the row equalizer,
we need to marginalize the $P\left ( \mathbf{\mathbf{y}_{k} \mid \widehat{y}}_{k} \right )$ over the two state bits $y(s_0) = y_1$ and 
$y(s_1) = y_2$. Similar marginalization is done for the
extrinsic probabilities $P\left ( \mathbf{y}_{k} \mid \mathbf{\widetilde{y}}_{k} \right )$. 
Based on the analysis in \cite{JZZ}, there is no 
need to subtract the input 16-valued LLRs from the output 16-valued LLRs 
when 
exchanging LLRs between row and column equalizers.

Details of the TDMR detector (without 2D-ISI) based on the BCJR algorithm are 
described in \cite{Carosino}. 
Here we describe a slight modification of the algorithm to interface the TDMR 
detector with the 2D-ISI equalizer. In particular, the gamma computation of 
the BCJR algorithm in the TDMR detector is implemented as follows: 
\begin{equation}
\begin{split}
&\gamma_{\mathbf{i}}\left ( \mathbf{\overline{y}}_{k},\mathbf{s'},\mathbf{s} \right )=\\
&\sum_{\mathbf{j}}[P'( y_{1}=j_{1},y_{2}=j_{2},y_{3}=j_{3},y_{4}=j_{4} \mid \mathbf{U}_k,S_{k},S_{k-1})\\
&\times P\left ( y_{1}=j_{1},y_{2}=j_{2},y_{3}=j_{3},y_{4}=j_{4}\right )] \times P(\mathbf{U}_k \mid \mathbf{s},\mathbf{s'})\\
&\times P\left ( \mathbf{s} \mid \mathbf{s'} \right )\times P\left ( \widetilde{u}_{k0}=i_{0} \right )\times P\left ( \widetilde{u}_{k1}=i_{1} \right ),
\label{eq:TDMR forward gamma}
\raisetag{10pt}
\end{split}
\end{equation}
where the $P(\mathbf{U}_{k} \mid \mathbf{s},\mathbf{s'})$ term is equal to 1/4, 
since the two coded bits in the input vector $\mathbf{U}_k$ are assumed to be independent 
of each other and of the grain states. The 
$P\left ( \mathbf{s} \mid \mathbf{s'} \right )$ term is the state transition 
probability computed from the grain connectivity table in \cite{Carosino}. 
The summation in the right-hand-side computes a weighted sum of the 
conditional probabilities in the original BCJR algorithm in \cite{Carosino} 
with extrinsic probabilities $P\left ( y_{1},y_{2},y_{3},y_{4}\right )$ from the 2D-ISI equalizer. The last two factors in (\ref{eq:TDMR forward gamma}) are 
probabilities of the
bits in input vector $\mathbf{U}_k = (u_{k0},u_{k1})$ computed from single valued 
LLRs from the SCCC decoder; they are computed as:
\[
%\begin{split}
P(\widetilde{u}_{km}=1)=\frac{e^{LLR_m}}{1+e^{LLR_m}}
%\label{eq:TDMR u1 compute}
%\end{split}
%\]
%\[
\text{ and }
%\begin{split}
P(\widetilde{u}_{km}=-1)=\frac{1}{1+e^{LLR_m}},
%\label{eq:TDMR u0 compute}
%\end{split}
\]
where $m \in \{0, 1\}$.

The TDMR detector's $\alpha$ and $\beta$ probabilities are updated
as in \cite{bcjr}, using the $\gamma$ probabilities
in (\ref{eq:TDMR forward gamma}).
Then the joint state and
input $\lambda$ probabilities at the $k$th step are
computed as 
\begin{equation}
\lambda_{k}^{\mathbf{i}}(\mathbf{s},\mathbf{s'}) = \alpha_{k-1}(S_{k-1}=\mathbf{s'})\gamma_{\mathbf{i}}(\mathbf{\overline{y}}_{k},\mathbf{s'},\mathbf{s})\beta_{k}(S_{k}=\mathbf{s}).
\label{eq:TDMR lambda}
\end{equation}
The LLRs sent from the TDMR detector to the SCCC decoder are given by 
\begin{equation}
%\begin{split}
LLR(u_{km})=\log\left ( \frac{\sum_{i_{\bar{m}}}\sum_{s}\sum_{s'}\lambda_{k}^{i_m=1,i_{\bar{m}}}(\mathbf{s},\mathbf{s'})}{\sum_{i_{\bar{m}}}\sum_{s}\sum_{s'}\lambda_{k}^{i_m=-1,i_{\bar{m}}}(\mathbf{s},\mathbf{s'})} \right ),
\label{eq:LLR to SCCC def}
%\end{split}
\end{equation}
where $m \in \{0, 1\}$ and $\bar{m} = (m + 1) \mod 2$.
The feedback probabilities sent from the TDMR detector to the 2D-ISI 
equalizer are computed as 
\begin{equation}
P(\mathbf{y}_{k})=\sum_{\mathbf{i},\mathbf{s},\mathbf{s'}}P(\mathbf{y}_{k} \mid \mathbf{U}=\mathbf{i},S_{k}=\mathbf{s},S_{k-1}=\mathbf{s'}) \lambda_{k}^{\mathbf{i}}(\mathbf{s},\mathbf{s'}),
\label{eq:prob feedback to 2DISI def}
\end{equation}
where the $\lambda$ probability is defined in (\ref{eq:TDMR lambda}),
and is used to weight the conditional probability when computing the 
probabilities for the vector $\mathbf{y}_k$. 
The LLRs sent to the 2D-ISI equalizer are given by 
\begin{equation}
\begin{split}
&LLR(\mathbf{y}_{k}=\mathbf{m}) = \\
&\log\left(\frac{\sum_{\mathbf{i},\mathbf{s},\mathbf{s'}}P(\mathbf{y}_{k}=\mathbf{m} \mid \mathbf{U},S_{k},S_{k-1})\times \lambda_{k}^{\mathbf{i}}(\mathbf{s},\mathbf{s'})}{\sum_{\mathbf{i},\mathbf{s},\mathbf{s'}}P(\mathbf{y}_{k}=\mathbf{-1} \mid \mathbf{U},S_{k},S_{k-1})\times \lambda_{k}^{\mathbf{i}}(\mathbf{s},\mathbf{s'})} \right ).
\label{eq:LLR feedback to 2DISI def}
\end{split}
\end{equation}

Because the extrinsic probability $P(y_1, y_2, y_3, y_4)$  
from the 2D-ISI to
the TDMR detector is
part of a weighted sum in the TDMR $\gamma$ probability 
(\ref{eq:TDMR forward gamma}), it cannot be factored out of the 
TDMR $\lambda$ probability (\ref{eq:TDMR lambda}), and thus
cannot form a separate term in the final LLR. Hence, 
we do not subtract the incoming 2D-ISI extrinsic information 
from the TDMR output LLRs in
(\ref{eq:LLR feedback to 2DISI def}) before sending them to
the 2D-ISI equalizer. Similarly, the marginalization of the incoming
probabilities $P\left (\mathbf{y}_{k} \mid \mathbf{\widehat{y}}_{k} \right )$ from the TDMR in (\ref{eq:2DISI gamma}) 
makes it impossible to retain these
probabilities as multiplicative factors in the final
$\lambda$ probabilities, making it inadvisable to perform
LLR subtraction before sending the 2D-ISI output LLRs to
the TDMR detector.   

\begin{figure}[htb]
  \begin{center}
  \includegraphics[width = 3.0in]{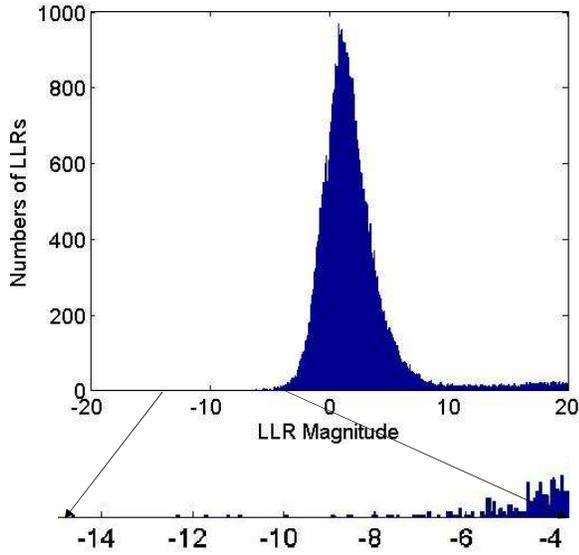}
\caption{Conditional distribution of TDMR output LLRs before weighting}
\label{fig:before}
  \end{center}
\end{figure}

\begin{figure}[htb]
  \begin{center}
  \includegraphics[width = 3.0in]{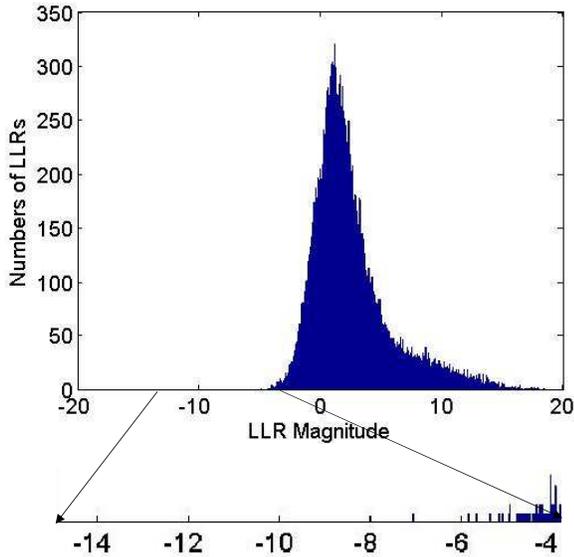}
\caption{Conditional distribution of TDMR output LLRs after weighting}
\label{fig:after}
  \end{center}
\end{figure}

To compensate for the lack of LLR subtraction in the 2D-ISI/TDMR loop,
a multiplicative weight factor $w \leq 1$ is applied 
to LLRs sent to the TDMR detector. This factor reduces the 
LLR magnitudes passed into the TDMR detector and helps
correct pixels that are incorrectly estimated by the 
2D-ISI equalizer, which have LLRs of large magnitude but with 
incorrect sign. 
For example, Fig.~\ref{fig:before} shows the histogram of output conditional 
LLRs of TDMR detector for the pixels corresponding to $+1$ codeword bits, 
before weight $w$ is applied, whereas Fig.~\ref{fig:after} shows the same 
LLRs after weighting. 
A positive LLR means the sign of original bit written on that pixel is 
successfully recovered by TDMR detector, whereas a negative LLR means that 
the result on that pixel remains incorrect. The weighting decreases large 
magnitude wrong LLRs as shown in the zoomed graph, which 
helps the SCCC decoder to correct the TDMR corrupted codeword.
The details of weight selection are described in the following 
section. 
%section~\ref{subsec:EXIT}.
%

\section{Optimization using EXIT Chart}
\label{sec:EXIT}
Proper design of the combined read process requires, among other things, 
specification of the weight $w$ mentioned in section \ref{sec:det/dec}, 
as well as the iteration schedule---how many 
inner loops of the SCCC decoder for each outer loop between TDMR detector/SCCC 
decoder and 2D-ISI equalizer/TDMR detector. This is a multi-parameter 
optimization problem; brute force search for an optimal choice of parameters 
by simulation of the complete system can be computationally expensive. 

This section describes the EXIT chart method to optimize the weight $w$ 
as well as the iteration schedule. 
EXIT charts were introduced by ten Brink \cite{tbrink}. They allow the 
performance of a system of concatenated detectors to be predicted based on 
input/output mutual information curves for individual constituent detectors, 
which can be computed relatively quickly. The mutual information $I_{A}$ 
between the input extrinsic information LLR to a given detector and the input 
codeword $X$ is computed; similarly the output mutual information $I_{E}$ 
between the output extrinsic information LLR from a given detector and the 
input codeword $X$ is computed. 
The mutual information $I_A$ and $I_E$ are defined as follows: 
\begin{equation}
\begin{split}
I_{A/E}=&\frac{1}{2}\sum_{x=-1,1}\int_{-\infty}^{+\infty}p_{A/E}(\xi \mid X=x)\\
&\times \log_{2}\frac{2p_{A/E}(\xi \mid X=x)}{p_{A/E}(\xi \mid X=-1)+p_{A/E}(\xi \mid X=1)}d\xi,
\raisetag{36pt}
\label{eq:EXIT def}
\end{split}
\end{equation}
where $p_{A/E}(\xi \mid X=x)$ is the experimental conditional PDF for input/output 
LLRs. The EXIT chart is obtained by plotting $I_{E}$ as a function of $I_{A}$ 
for both TDMR detector and SCCC decoder on the same set of axes, using the 
horizontal axis as $I_{A}$ for TDMR detector and the vertical axis as $I_{A}$ 
for SCCC decoder. Thus a pair of $I_A$ vs $I_E$ curves are obtained for  
different values of the parameter to be optimized (e.g., weight $w$). 
The optimal value of the parameter is one that results in the corresponding 
pair of $I_A$ vs $I_E$ curves being close to each other, without touching 
or intersecting. In this paper, we did not jointly optimize the weight $w$ 
and the iteration schedule. Instead, we optimized the weight parameter 
first, for a fixed iteration schedule. Next, we fixed the weight $w$ found 
in the first step and optimized the iteration schedule. 

\begin{figure}[tbh]
  \begin{center}
  \includegraphics[width=3.5in]{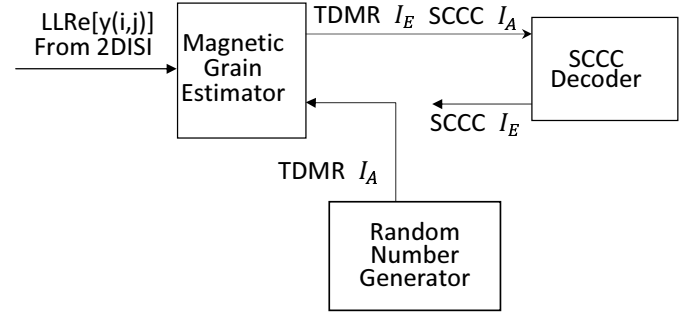}
\caption{LLR injection block diagram}
\label{fig:EXIT block}
  \end{center}
\end{figure}

For EXIT chart optimization, the goal is to select a set of parameter values that 
results in $I_A$ vs $I_E$ curves satisfying the requirement of not touching 
with the highest code rate. The EXIT chart optimization in this section is based on the grain image with $P_{2}=0.2$.
In actual experiments, we noticed that the input mutual information varies 
over only a small range of values. This makes it difficult to generate a 
complete set of EXIT chart curves. 
To ameliorate this problem, we simulate input LLRs using a random number 
generator, whose distribution is close to that of the observed histogram 
of the input LLR. The random numbers are then injected into TDMR detector as 
shown in Fig.~\ref{fig:EXIT block}.

\begin{figure}[tbh]
  %\begin{center}
%  \epsfig{figure=fit.eps,width=3.5in}
  \includegraphics[width=2.8in]{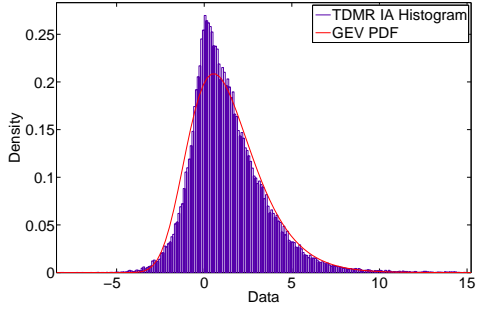}
\caption{Fitting a GEV distribution to an experimental LLR histogram;
the GEV parameters are $(k,\sigma,\mu) = (-0.06366,1.766,0.4258)$.}
\label{fig:GEV}
  %\end{center}
\end{figure}

Experimentally, we observed that the general extreme value (GEV) distribution 
was a good fit to the observed TDMR $I_A$ LLR histogram. 
The probability density function $f(x)$ of the GEV distribution is given by
\begin{equation}
\begin{split}
f(x)=\frac{1}{\sigma }t(x)^{k+1}e^{-t(x)},
\ \text{where}\quad\quad\quad\quad\ &\\
t(x)=\left\{\begin{matrix}\left ( 1+\left ( \frac{x-\mu}{\sigma} \right )k \right )^{-\frac{1}{k}},&\quad k\neq 0
\\e^{-\frac{x-\mu}{\sigma}},&\quad k=0.
\\
\end{matrix}\right.
\label{eq:GEV PDF}
\end{split}
\end{equation}
Fig.~\ref{fig:GEV} shows the fit of the GEV distribution to a typical 
experimental TDMR $I_A$ LLR histogram. 

\begin{figure}[tb]
\setlength{\abovecaptionskip}{0pt}
\setlength{\belowcaptionskip}{0pt}
\begin{center}
\includegraphics[width=3.5in]{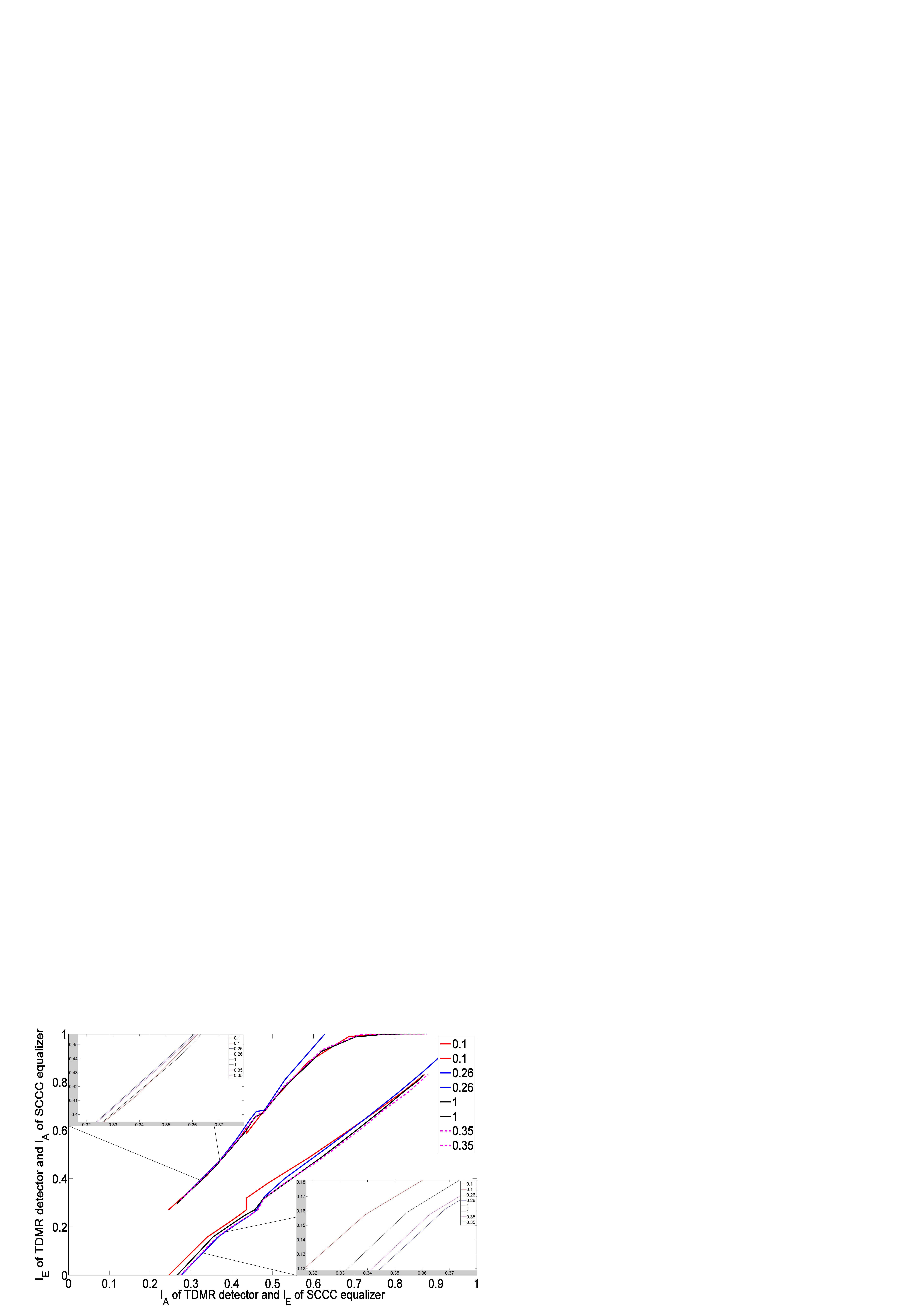}
\caption{EXIT chart with different weights $w$}
\label{fig:EXIT}
\end{center}
\end{figure}

The input mutual information can now be varied by appropriately changing the 
parameters of GEV. 
Based on experimental LLR histograms, the parameter $k$ of the GEV 
distribution varies very little when the weight $w$ or iteration schedule varies. 
%compared to parameters $\mu$ and $\sigma$. 
Therefore we fixed the parameter value of $k = -0.0658773$---based on experiments---in our 
random number generator. 
The parameters $\mu$ and $\sigma$ we observed to satisfy (approximately) the 
linear relation $\sigma = 2.7170 \mu + 0.5586$. 
However, we set an upper limit for the parameter $\sigma$ 
(4.37 in our experiments) so that the width of the GEV distribution does 
not grow too large; i.e., we set 
$\sigma = \min (2.7170 \mu + 0.5586, 4.37)$ and vary $\mu$ to simulate 
different input mutual information values to generate the $I_A$ vs $I_E$ 
curves. Fig.~\ref{fig:EXIT} depicts these curves for different values of 
the weight $w$. Two inner loops of the SCCC decoder were used for this part 
of the experiment.
%
%%%%
%Jiyang: What was the iteration schedule for this set of curves?
%%%%

The EXIT chart curves for different weights are fairly close to each other. 
We numerically computed the distance between pairs of $I_A$ vs $I_E$ curves 
to determine the weight $w$ that yields the maximum distance.
Typically, the experimental SCCC input mutual information $I_{A}$ never 
rises much higher than 
0.4. Therefore, we used the distance between the second points 
on the two 
curves as the distance metric. (Out of 13 points on each curve, 
the second point always has SCCC $I_{A} \approx 0.4$.)
Based on this metric, we determined that the 
weight $w = 0.26$ results in the farthest apart
pair of $I_A$ vs $I_E$ curves. 

Next, we fixed the weight $w = 0.26$ and obtained a set of EXIT chart curves by varying 
the iteration schedule---number of inner loops in the SCCC decoder. Again, we 
numerically computed the distance between pairs of $I_A$ vs $I_E$ curves 
to find the iteration schedule that yields the maximum distance. 
Based on this metric, we determined that ten inner loops of the SCCC decoder 
gives the farthest apart pair of $I_A$ vs $I_E$ curves. 

%%%%
%Jiang: Need a little more explanation here.
%Was weight w set at 0.26 here and the SCCC inner loops varied?
%%%%
%Same method is employed on optimization of SCCC iteration schedule. 
%The result of optimized number of iterations in SCCC decoder is 10 iterations.

\section{Simulation Results}
\label{sec:sims}

%%%%
%Jiyang: Were you going to add another subsection later?
%If not, no need for a subsection here.
%%%%

%\subsection{TDMR/SCCC Interface and Iteration Schedule}
%\label {subsec:det_dec}
This section presents Monte Carlo simulation results for the system 
described in section \ref{sec:det/dec}, and compares 
its performance to previously published results by Pan and Ryan 
et.\ al.\ \cite{LuPan-jour}. As discussed in section \ref{sec:EXIT}, we 
set the weight $w = 0.26$, and run ten inner loops of the SCCC decoder for 
each outer loop of the TDMR detector/SCCC decoder. Three iterations of row/column 2D-ISI equalizer loop and six iterations of the 
TDMR detector/SCCC decoder loop are done for each iteration of the 
2D-ISI equalizer/TDMR detector loop; this outer loop schedule was
optimized by multiple simulation runs. 

The simulations employ multiple TDMR images; each image 
is written by the SCCC codeword corresponding
to 32,768 randomly generated equiprobable information bits. The encoder 
described in section \ref{sec:model_arch} is used. The coded bits 
are arranged into a rectangular image with 512 columns. 
The number of rows depends on the total number of codeword bits, which 
depends on the code rate. A nominal code rate of 1/4 gives 
256 rows. The code rate is adjusted by suitable puncturing or 
repetition of the coded bits. 
2D-ISI with the $2 \times 2$ averaging mask and AWGN is imposed on 
the image as described in section \ref{sec:model_arch}. 
The simulations are performed for 
two different SNR values of 9 dB and 10 dB, computed as in (\ref{eq:SNR def}). 
For each value of the grain probability $P_{2}$, we adjust the code rate 
to achieve a decoded BER of $10^{-5}$. This is ensured 
by decoding at least 100 codeword blocks of 32,768 information bits each, 
with at most 32 errors. 
The maximum code rate achieving this BER is recorded for 
each value of $P_{2}$.

\begin{figure*}[t!]
  \begin{center}
\includegraphics[width=5.9in]{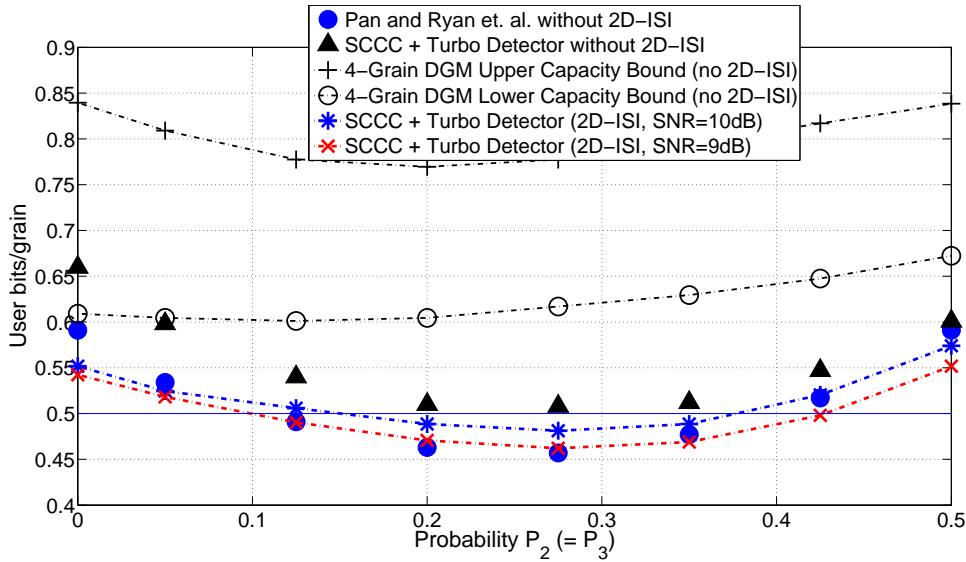}
\caption{Simulation results depicting the performance of the proposed 
combined 2D-ISI equalizer, TDMR detector, and SCCC decoder.}
\label{fig:simulation}
  \end{center}
\end{figure*}

In Fig.~\ref{fig:simulation}, the horizontal axis represents the value of 
$P_{2}$ and the vertical axis represents the value of user bits per grain, 
which is twice the code rate. The triangle marker represents the performance 
of the system with no 2D-ISI (or additive noise) and is reproduced from 
\cite{Carosino}. 
%The triangles represent the code rate achieved by TDMR detector and SCCC decoder with iterations. 
The asterisk marker with dash-dotted line is the corresponding result for the 
system with 2D-ISI and additive noise with 10dB SNR. Results for 9 dB SNR are 
depicted with an 'X' and dash-dotted line. Lower and upper bounds 
(from \cite{Kavcic}) on the
channel capacity of the 4-grain DGM TDMR channel (without 2D-ISI and AWGN) 
are shown with the open-circle and '+' markers, respectively.

\begin{table*}[t!]\normalsize
\caption{Rate penalty with respect to system without 2D-ISI.}
\label{table:result}
\begin{center}
\begin{tabular}{|c|c|c|c|c|c|c|c|}
\hline

\multirow{2}{*}{$P_2$} &
\multicolumn{3}{c|}{\tabincell{c}{SCCC/TDMR system\\ with \\2D-ISI and SNR=10dB}} &
\multicolumn{3}{c|}{\tabincell{c}{SCCC/TDMR system\\ with \\2D-ISI and SNR=9dB}} \\
 
\cline{2-7}
 & \tabincell{c}{Rate penalty\\ w.r.t system \\without 2D-ISI} & \tabincell{c}{Rate gain\\ w.r.t result \\in \cite{LuPan-jour}} & \tabincell{c}{SNR/info bit} & \tabincell{c}{Rate penalty\\ w.r.t system \\without 2D-ISI} & \tabincell{c}{Rate gain\\ w.r.t result \\in \cite{LuPan-jour}} & \tabincell{c}{SNR/info bit} \\

\hline
 
0 & 16.4\% & -6.6\% & 15.59dB & 17.8\% & -8.2\% & 14.67dB \\
\hline
0.05 & 12.3\% & -1.7\%& 15.81dB & 13.4\% & -2.9\% & 14.86dB \\
\hline
0.125 & 6.3\% & 3.0\%& 15.97dB & 9.2\% & -0.1\% & 15.10dB \\
\hline
0.2 & 4.2\% & 5.5\%& 16.12dB & 7.7\% & 1.6\% & 15.28dB \\
\hline
0.275 & 5.3\% & 5.3\%& 16.19dB & 9.0\% & 1.1\% & 15.36dB \\
\hline
0.35 & 4.6\% & 2.4\%& 16.12dB & 8.4\% & -1.7\% & 15.30dB \\
\hline
0.425 & 4.9\% & 0.6\%& 15.85dB & 8.9\% & -3.7\% & 15.04dB \\
\hline
0.5 & 4.5\% & -2.8\%& 15.42dB & 8.2\% & -6.6\% & 14.59dB \\
\hline

\end{tabular}
\end{center}
\end{table*}

Table~\ref{table:result} shows the code rate penalty (as a percentage) 
of the system with 2D-ISI and AWGN with respect to the TDMR/SCCC system 
without 2D-ISI and AWGN, along with the code rate gain (as a percentage) with respect to the system proposed by Pan and Ryan 
et.\ al.\ \cite{LuPan-jour}. The SNR (\ref{eq:SNR def}) used in
Fig.~\ref{fig:simulation} is per 
codeword bit. Table~\ref{table:result} also shows the SNR per information
bit (taking into account the code rate) for both the 9 and 10 dB curves 
in Fig.~\ref{fig:simulation}.
Compared to the result of pure TDMR case, the performance 
degradation due to 2D-ISI and additive noise is obvious and expected. 
At the left end, for $P_{2}$ values of 0 and 0.125, the decrease in code rate 
performance due to 2D-ISI and noise is relatively large---12\% to 16\% at 
10 dB and 13\% to 18\% at 9 dB. 
%At the right end, for $P_{2}$ value of 0.5, modeling difficulty is described in [Michael's paper]. 
For the central part of $P_{2}$ axis, where the model achieves a closer approximation of real magnetic grain channel, our system with 2D-ISI and 
additive noise outperforms the results published by Pan, Ryan et.\ al.\ 
---shown with filled circle marker---with no 2D-ISI or additive noise. 
At 10 dB SNR, the minimum density of 0.48 user bits per grain achieved by our 
combined iterative detector/equalizer/decoder occurs at a $P_{2}$ value of 0.275,
and corresponds to an on-disk areal density of about 9.6 Terabits/$\mathrm{in^2}$,
under the typically assumed media grain density of 20 Teragrains/$\mathrm{in^2}$.
Thus, our simulation results support the feasibility of TDMR at densities
of about 10 Terabits/$\mathrm{in^2}$, as proposed in \cite{Roger}.
\section{Conclusion}
\label{sec:conc}

This paper introduces a system for iterative detection, equalization, and 
decoding of two-dimensional magnetic recording channels in the presence of 2D-ISI and 
AWGN. It consists of a 2D-ISI equalizer, a TDMR detector based on a rectangular
4-grain DGM, and a SCCC decoder. 
The equalization algorithm uses joint extrinsic information to tackle 
the 2D-ISI and AWGN in a TDMR corrupted grain image. Methods for incorporating
extrinsic information passed between the 2D-ISI and TDMR modules are presented. 
%Modification is done for this equalization algorithm and TDMR detection BCJR algorithm according to the independence assumption on the extrinsic information exchanged among detectors. 
Optimal settings for some of the parameters of the system (e.g., weight 
factor applied on LLRs sent from 2D-ISI equalizer to TDMR detector, and the 
iteration schedule in the SCCC decoder) are obtained using an EXIT chart method. 
Performance of the optimized system is presented and compared to that 
of previously published results for equivalent TDMR channels without 2D-ISI. For the most 
part, the proposed system with 2D-ISI outperforms these previously published results.
The presented simulations suggest that practical
TDMR systems can achieve the density of 10 Terabits/$\mathrm{in^2}$ predicted
in earlier publications; however, it must be noted that the rectangular DGM is somewhat
simplistic compared to actual grain shapes. 
Future work will consider more realistic grain models and higher performance codes like LDPC codes. The capacity of the combined TDMR/2D-ISI channel
described by (\ref{eq:y_u_relation_new}) and (\ref{eq:2DISIequ}) remains
an open problem.

\begin{comment}
Future extension of this work includes further optimization of the system by 
modifying the algorithm in TDMR detector---computing $\gamma$ probability 
for each possible $\mathbf{y}$ vector pattern instead of averaging the 
extrinsic information obtained from 2D-ISI equalizer. 
Further performance improvement is expected if the modification is implemented.
\end{comment}

\section*{Acknowledgment}

This work was supported by NSF grants CCF-1218885 and CCF-0635390.
The authors also wish to acknowledge useful discussions with
Dr. Roger Wood of Hitachi Global Storage Technologies, San Jose, CA.

\bibliographystyle{IEEEtran}
\bibliography{tdmr_ref,zzjour_chen_rev2_dir2disi-refer_corrected}

\begin{thebibliography}{1}
\providecommand{\url}[1]{#1}
\csname url@samestyle\endcsname
\providecommand{\newblock}{\relax}
\providecommand{\bibinfo}[2]{#2}
\providecommand{\BIBentrySTDinterwordspacing}{\spaceskip=0pt\relax}
\providecommand{\BIBentryALTinterwordstretchfactor}{4}
\providecommand{\BIBentryALTinterwordspacing}{\spaceskip=\fontdimen2\font plus
\BIBentryALTinterwordstretchfactor\fontdimen3\font minus
  \fontdimen4\font\relax}
\providecommand{\BIBforeignlanguage}[2]{{%
\expandafter\ifx\csname l@#1\endcsname\relax
\typeout{** WARNING: IEEEtran.bst: No hyphenation pattern has been}%
\typeout{** loaded for the language `#1'. Using the pattern for}%
\typeout{** the default language instead.}%
\else
\language=\csname l@#1\endcsname
\fi
#2}}
\providecommand{\BIBdecl}{\relax}
\BIBdecl

\bibitem{Roger}
R.~Wood, M.~Williams, A.~Kavcic, and J.~Miles, ``The feasibility of magnetic
  recording at 10 terabits per square inch on conventional media,'' \emph{IEEE
  Trans. Magnetics}, vol.~45, no.~2, pp. 917--923, Feb. 2009.

\bibitem{Kavcic}
A.~Kavcic, X.~Huang, B.~Vasic, W.~Ryan, and M.~F. Erden, ``Channel modeling and
  capacity bounds for two-dimensional magnetic recording,'' \emph{IEEE Trans.
  Magnetics}, vol.~46, no.~3, pp. 812--818, Mar. 2010.

\bibitem{LuPan-jour}
L.~Pan, W.~E. Ryan, R.~Wood, and B.~Vasic, ``Coding and detection for
  rectangular-grain {TDMR} models,'' \emph{IEEE Trans. Magnetics}, vol.~47,
  no.~6, pp. 1705--1711, June 2011.

\bibitem{bcjr}
L.~R. Bahl, J.~Cocke, F.~Jelinek, and J.~Raviv, ``Optimal decoding of linear
  codes for minimizing symbol error rate,'' \emph{IEEE Transactions on
  Information Theory}, vol.~20, pp. 284--287, March 1974.

\bibitem{Carosino}
M.~Carosino, Y.~Chen, B.~J. Belzer, K.~Sivakumar, J.~Murray, and P.~Wettin,
  ``Iterative detection and decoding for the four-rectangular-grain {TDMR}
  model,'' in \emph{Proceedings of the Allerton Conference on Communication,
  Control, and Computing (accepted, to appear)}, 2013, also available at
  http://arxiv.org/abs/1309.7518.

\bibitem{JRC}
Y.~Chen, B.~J. Belzer, and K.~Sivakumar, ``Iterative row-column soft-decision
  feedback algorithm using joint extrinsic information for two-dimensional
  intersymbol interference,'' in \emph{Proceedings of the 44th Annual
  Conference on Information Sciences and Systems (CISS 2010)}, Princeton, NJ,
  March 2010, pp. 1--6.

\bibitem{JZZ}
------, ``Iterative soft-decision feedback zigzag algorithm using joint
  extrinsic information for two-dimensional intersymbol interference,'' in
  \emph{Proceedings of the 45th Annual Conference on Information Sciences and
  Systems (CISS 2011)}, Baltimore, MD, March 2011, pp. 1--6.

\bibitem{tbrink}
S.~ten Brink, ``Convergence behavior of iteratively decoded parallel
  concatenated codes,'' \emph{IEEE Trans. Commun.}, vol.~49, no.~10, pp.
  1727--1737, Oct. 2001.

\end{thebibliography}

\end{document}